\documentclass{elsart}
\usepackage{epsfig}
\usepackage{amssymb}
\begin{document}
\vglue 1.5 true cm
\begin{frontmatter}
\title{Quantum interference of $\rho^0$- and $\omega$-mesons \\
 in the $\pi\, N \rightarrow e^+e^- N$ reaction}

\author{Matthias F. M. Lutz}
\address{GSI, Planckstrasse 1,  D-64291 Darmstadt, Germany\\
Institut f\"ur Kernphysik, TU Darmstadt,
   D-64289 Darmstadt, Germany}
\author{Bengt Friman}
\address{GSI, Planckstrasse 1,  D-64291 Darmstadt, Germany \\
Institut f\"ur Kernphysik, TU Darmstadt,
   D-64289 Darmstadt, Germany}
\author{Madeleine Soyeur}
\address{D\'{e}partement d'Astrophysique, de Physique des Particules,\\
de Physique Nucl\'{e}aire et de l'Instrumentation Associ\'{e}e,\\
Service de Physique Nucl\'{e}aire,
CEA/Saclay,
\\F-91191 Gif-sur-Yvette Cedex, France}

\begin{abstract}

   The study of the $\pi N\rightarrow \rho^0 N$ and
$\pi N \rightarrow \omega N$ amplitudes below and close to the
vector meson production threshold ($1.4<\sqrt s <1.8$ GeV)
reveals a rich structure arising from the presence of
baryon resonances in this energy range. These resonances are reflected in the
interference pattern of the $e^+e^-$ decays
of the $\rho^0$- and $\omega$-mesons produced in
$\pi^- p$ and $\pi^+ n$ reactions. We discuss
the shape and magnitude of the $\rho^0$-$\omega$ interference
in the $\pi^-p \rightarrow e^+e^- n$ and $\pi^+n \rightarrow e^+e^- p$
reaction cross sections
as functions of the total center of mass energy $\sqrt s$. We find contrasted results:
the interference is largely destructive for
the $\pi^-p \rightarrow e^+e^- n$ cross section but
constructive for the $\pi^+n \rightarrow e^+e^- p$ cross section.
An experimental study of these reactions would provide significant constraints
on the coupling of vector meson-nucleon channels to low-lying baryon resonances.

\vskip 0.3truecm

\noindent
{\it Key words}: Vector meson production; Baryon  resonances; Dileptons;
Quantum interference

\noindent
{\it PACS:} 13.20; 13.75.G; 14.20.G
\end{abstract}

\end{frontmatter}
\newpage

\section{Introduction}

Baryon resonances have been most extensively studied in partial-wave analyses
of pion-nucleon elastic scattering data.
Fairly accurate values of their masses, half-widths and pion-nucleon
partial decay widths have been extracted up to pion-nucleon center
of mass energies of about 2 GeV \cite{Arndt1,Arndt2}.
Isobar analyses of inelastic data ($\pi N \rightarrow \pi \pi N$)
make it possible to study meson-nucleon final states other than
$\pi N$, such as $\rho N$ and $\pi N^*$
(where $N^*$ denotes a low-lying baryon resonance which couples strongly
to the $\pi N$ channel) \cite{Manley1,Vrana}. The
$\pi N \rightarrow \pi \pi N$ data are known with much less statistical
accuracy than the pion-nucleon elastic scattering data. For some partial waves,
specific final states besides the $\pi N$ and $\pi \pi N$ channels
have to be included in the analyses to satisfy unitarity. The
$\eta N$ channel in particular is needed at low energy for a
proper description of the S$_{11}$ wave \cite{Manley1,Vrana}.
Similarly, at higher energies, the $\omega N$ and $K \Lambda$
channels play a role \cite{Manley1,Vrana}.

The study of the $\pi^-p \rightarrow e^+e^- n$ and $\pi^+ n \rightarrow e^+e^- p$ processes described
in this work aims at gaining understanding of the
$\pi N \rightarrow \rho^0 N$ and $\pi N \rightarrow \omega N$
scattering amplitudes for center of mass ener\-gies close and below the
vector meson production threshold ($1.5<\sqrt s <1.8$ GeV).
There are well-known baryon resonances in this energy range,
which contribute
to the $\pi^-p \rightarrow e^+e^- n$ and $\pi^+ n \rightarrow e^+e^- p$
scattering amplitudes through their coupling to the $\rho^0 N$ and $\omega N$ channels.
These amplitudes involve in addition significant non-resonant processes.

Phenomenological constraints  on the $\rho N N^*$ and $\omega N N^*$
coupling strengths are useful both for baryon structure studies
and for building models of vector meson propagation in the
nuclear medium. On the one hand,
it is of interest to compare the $\rho N N^*$ and $\omega N N^*$
coupling strengths entering the description of
the  $\pi^-p \rightarrow e^+e^- n$ and $\pi^+ n \rightarrow e^+e^- p$
cross sections to quark model predictions for the corresponding
quantities. In particular, the ${e^+e^-}$ channel offers the possibility
to study the coupling of low-lying baryon resonances to the
$\omega N$ channel below threshold.
On the other hand,
the $\rho N N^*$ and $\omega N N^*$ coupling strengths
determine the contribution
of resonance-hole states to the $\rho^0$- and $\omega$-meson propagator
in nuclear matter. It has been suggested that the spectral distribution of those mesons
at high baryon density could be
interpreted as a signal of chiral symmetry restoration in the nuclear medium \cite{Brown1}
and hence be sensitive to the nonperturbative
structure of Quantum Chromodynamics.
The spectral functions of vector mesons in the nuclear medium
should be reflected in the spectra of lepton pairs
produced in photon- and hadron-nucleus reactions as well as in ultra-relativistic heavy ion collisions
\cite{Rapp}. A proper understanding of the
$\pi^-p \rightarrow e^+e^- n$ and $\pi^+ n \rightarrow e^+e^- p$
 reactions appears as a first and necessary step
towards a detailed interpretation of the production of lepton pairs off nuclei
induced by charged pions.
 Because of the very large width of $\rho$-mesons in nuclei,
the production of $e^+e^-$ pairs in pion-nucleus interactions is expected
to be mostly sensitive to the in-medium propagation of $\omega$-mesons
 for $e^+e^-$ pair invariant masses close to
vector meson masses \cite{Schoen,Effenberger}.
Both the $\pi N\rightarrow e^+e^- N$
and $\pi A \rightarrow {e^+e^-} X$  cross sections  could be measured at GSI-Darmstadt
with the HADES detector \cite{Schoen,HADES}.

The exclusive observation of neutral vector mesons through their $e^+e^-$
decay presents definite advantages over their observation
through final states invol\-ving pions.
Firstly, there are no competing processes, such as $\pi \Delta$ production leading to the same final state and
impairing the identification of the $\rho$-meson
in the $\pi \pi N$ channel. Se\-condly, both the $\rho^0$- and
$\omega$-mesons decay into the $e^+e^-$ channel.
This leads to a quantum interference pattern which is expected to reflect
the structure and relative sign of the $\pi N \rightarrow \rho^0 N$
and $\pi N \rightarrow \omega N$ scattering amplitudes.
Such an interference in the $e^+e^-$ channel,
observed in the photoproduction of
$\rho^0$- and $\omega$-mesons at higher energies \cite{Alvensleben}, has
proved very useful in establishing the similarity of
the $\gamma p \rightarrow \rho^0 p$ and the
$\gamma p \rightarrow \omega p$ processes
and the diffractive nature of these reactions for incident photon energies
of a few GeV.

The $\pi N \rightarrow e^+e^- N$ cross section is connected to the
$\pi N \rightarrow \rho^0 N$ and $\pi N\rightarrow \omega N$
scattering amplitudes by the Vector Meson Dominance assumption \cite{Sakurai,Kroll}.
In this picture, the produced  $\rho^0$- or
$\omega$-meson is converted into an intermediate time-like photon
which subsequently materializes into an $e^+e^-$ pair.
The dynamics of the $\pi N \rightarrow e^+e^- N$, $\gamma N \rightarrow \pi N$ and
$e^- N \rightarrow e^- \pi N$ processes is expected to be similar
and dominated by the couplings of baryon resonances to the pion and to vector fields.
In that sense,
the study of the $\pi N \rightarrow e^+e^- N$ reaction
complements programs devoted to the photo- and electroproduction
of baryon resonances. There is presently a large activity
in this field at ELSA
\cite{ELSA}, MAMI \cite{MAMI} and the Jefferson Laboratory
\cite{JLAB}.

To discuss the $\pi N\rightarrow e^+e^- N$ reaction, we use
the $\pi N \rightarrow \rho^0 N$ and $\pi N \rightarrow \omega N$
amplitudes obtained in the recent unitary coupled-channel
model of Ref. \cite{Lutz1}. This is a relativistic approach
where, in contrast to
isobar analyses of $\pi N$ scattering, the s- and d-wave pion-nucleon
resonances are generated dynamically starting from an effective field theory of
meson-nucleon scattering \cite{Lutz1}. We present the model
and its predictions for the $\pi N \rightarrow \rho^0 N$ and $\pi N
\rightarrow \omega N$ amplitudes in Section 2. The calculation of the
$\pi^-p \rightarrow e^+e^- n$ and
$\pi^+n \rightarrow e^+e^- p$
cross sections in the
Vector Meson Dominance model is outlined in Section 3. Our numerical results
for these cross sections
are displayed in Section 4.
We show how the $\rho^0-\omega$ quantum interference pattern in the
$e^+e^-$ spectrum occurs in both isospin channels and evolves as function of the total
pion-nucleon center of mass energy in the interval
($1.5<\sqrt s <1.8$ GeV).
Concluding remarks are given in Section 5.

\newpage

\section{The  $\pi N \rightarrow \rho^0 N$ and $\pi N
\rightarrow \omega N$ amplitudes close to the vector
meson production threshold}

We describe the $\pi N \rightarrow e^+e^- N$ reaction for
$e^+e^-$ pair invariant masses ran\-ging from $\sim$0.4 to $\sim$0.8 GeV.
The exclusive measurement of the $e^+e^- N$ outgoing channel
ensures that the $e^+e^-$ pair comes from the decay of a time-like
photon. In this respect, the identification of the $e^+e^-$ decay of
vector mesons is easier in the $\pi N \rightarrow e^+e^- N$
process than in the photoproduction reaction $\gamma N \rightarrow e^+e^- N$,
where it interferes with Bethe-Heitler pair production.
Assuming Vector Meson Dominance for the electromagnetic current \cite{Kroll},  the
$\pi N \rightarrow \rho^0 N$ and $\pi N
\rightarrow \omega N$ amplitudes are the basic
quantities entering the calculation of the $\pi N \rightarrow e^+e^- N$
cross section. At pion-nucleon center of mass energies close to the
$\rho^0$- and $\omega$-meson production threshold
the kinetic energy of the particles in the entrance channel is large.
Hence a relativistic description of the $\pi N \rightarrow \rho^0 N$ and $\pi N
\rightarrow \omega N$ amplitudes is preferable.

We study the $\pi^-p \rightarrow e^+e^- n$ and $\pi^+n \rightarrow e^+e^- p$
reactions in the framework
of a recent relativistic and unitary coupled-channel approach to
meson-nucleon scattering \cite{Lutz1}.
The available data on pion-nucleon elastic and inelastic scatte\-ring and
on meson photoproduction off nucleon targets
are fitted in the energy window $1.4<\sqrt s <1.8$ GeV, using an effective
Lagrangian with quasi-local two-body meson-baryon interactions
and a generalized form of Vector Meson Dominance to describe the coupling
of vector mesons to real photons.
The scheme involves the $\pi N$, $\pi \Delta$, $\rho N$,
$\omega N$, $K \Lambda$, $K \Sigma$ and $\eta N$ hadronic channels, whose
relevance in this energy range has been highlighted
by earlier non-relativistic analyses \cite{Manley1,Vrana}.
The coupling cons\-tants entering the effective Lagrangian are parameters
which are adjusted to reproduce the data. In view of the kinematics,
only s-wave scattering in the $\rho N$ and $\omega N$ channels is included,
restricting $\pi N$ and $\pi \Delta$ scattering to s- and d-waves.
The pion-nucleon resonances in the S$_{11}$, S$_{31}$, D$_{13}$ and D$_{33}$
partial waves are generated dynamically by solving Bethe-Salpeter
equations \cite{Lutz1}.

The model provides a good fit to the
$\pi^-p \rightarrow \rho^0 n $, $\pi^- p \rightarrow \omega n$
and $\gamma p\rightarrow \rho^0 p$
cross sections for $\sqrt s \leq 1.75$ GeV, where enough data close to threshold are available for
comparison \cite{Lutz1}.
Above threshold, where higher partial waves should become important (in channels where
pion-exchange effects are expected to be large), it underestimates
significantly the $\pi^-p \rightarrow \rho^0 n $ and the $\gamma p\rightarrow \omega p$ cross sections.
The model reproduces satisfactorily the $\pi N$ scattering data (phase shifts and inelasticities)
and pion photoproduction multipole amplitudes below the vector meson production threshold
\cite{Lutz1}. Our approach is
therefore appropriate to values of $\sqrt s \leq 1.75$ GeV.
In the $\rho^0 N$- and $\omega N$-channels,
the restriction to s-wave scattering means that the model applies
to situations where the vector meson is basically at rest with
respect to the scattered nucleon $(\sqrt s \simeq M_N + M_V)$, where
$M_N$ and $M_V$ denote the nucleon and the vector-meson masses respectively.
This assumption implies that the range of validity of the present
calculation is limited to $e^+e^-$ pairs with invariant masses
$m_{e^+e^-}$ close to $(\sqrt s - M_N)$.

The $\pi N \rightarrow \rho N$ and $\pi N \rightarrow \omega N$ amplitudes are represented
diagrammatically in Fig. 1.
The $\pi N \rightarrow \rho N$ amplitude has isospin 1/2
and isospin 3/2 components while the  $\pi N \rightarrow \omega N$ amplitude
selects the isospin 1/2 channel. Both amplitudes have spin 1/2 and spin 3/2
parts.

\noindent
\begin{figure}[h]
\noindent
\begin{center}
\mbox{\epsfig{file=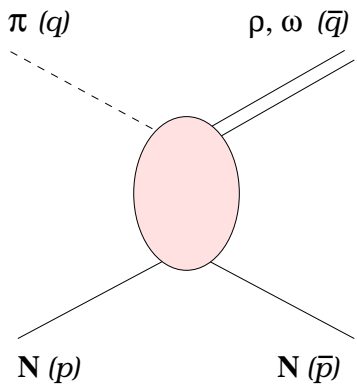, height=5cm}}
\end{center}
\vskip 0.3 true cm
\caption{The $\pi N  \rightarrow \rho (\omega) N$ amplitude.}
\label{f1}
\end{figure}

The invariant transition matrix elements for the $\pi N \rightarrow \rho N$
and $\pi N \rightarrow \omega N$ reactions are given by

\begin{eqnarray}
\langle \rho^j (\overline{q})\,N(\overline{p})|\,{\mathcal T}\,| \pi^i(q)\, N(p)\rangle & \nonumber \\
= \,(2\pi)^4\,  \delta^4 (q+ &p- \overline{q}- \overline{p})\, \overline u(\overline{p})\, \epsilon^\mu (\overline{q})
\,T^{ij}_{(\pi N\rightarrow \rho N)\,\mu}\,u(p),
\label{eq:e1}
\end{eqnarray}
\begin{eqnarray}
\langle \omega (\overline{q})\,N(\overline{p})|\,{\mathcal T}\,| \pi^i(q)\, N(p)\rangle & \nonumber \\
= \,(2\pi)^4\,  \delta^4 (q+ &p- \overline{q}- \overline{p})\, \overline u(\overline{p})\, \epsilon^\mu (\overline{q})
\,T^i_{(\pi N\rightarrow \omega N)\,\mu}\,u(p),
\label{eq:e2}
\end{eqnarray}

\noindent
where $T^{ij}_{(\pi N\rightarrow \rho N)\,\mu}$ and
$T^{i}_{(\pi N\rightarrow \omega N)\,\mu}$ are functions of the three kinematic variables
$w=p+q=\overline{p}+\overline{q}$ ($\sqrt {w^2}$ = $\sqrt s$), $q$ and
$\overline{q}$.

These scattering amplitudes can be decomposed into isospin invariant components
as \cite{Lutz2}
\begin{eqnarray}
T^{ij}_{(\pi N\rightarrow \rho N)\, \mu} (\overline{q},q;w)
\, = \sum_{I}
T^{(I)}_{(\pi N \rightarrow \rho N)\,\mu}(\overline{q},q;w)\, P_{(\rho)}^{(I)\,ij},
\label{eq:e5}
\end{eqnarray}
\begin{eqnarray}
T^{i}_{(\pi N\rightarrow \omega N)\, \mu} (\overline{q},q;w)
\, = \sum_{I}
T^{(I)}_{(\pi N \rightarrow \omega N)\, \mu}\,(\overline{q},q;w) P_{(\omega)}^{(I)\,i},
\label{eq:e6}
\end{eqnarray}

in which the isospin projectors are given for the $\pi N\rightarrow \rho N$
transition by

\begin{eqnarray}
P_{(\rho)}^{(I=\frac {1} {2}) \,i,j} \, = \, \frac {1} {3} \, \tau^i \tau^j,
\label{eq:e7}
\end{eqnarray}

\begin{eqnarray}
P_{(\rho)}^{(I=\frac {3} {2}) \,i,j} \, = \, \delta^{ij} - \frac {1} {3} \,
 \tau^i \tau^j,
\label{eq:e8}
\end{eqnarray}

and for the $\pi N\rightarrow \omega N$ transition by

\begin{eqnarray}
P_{(\omega)}^{(I=\frac {1} {2}) \, i} \, = \, \frac {1} {\sqrt 3} \, \tau^i.
\label{eq:e9}
\end{eqnarray}

The isospin invariant amplitudes can be expanded further into
components of total angular momentum using the relativistic
projection operators introduced in Ref. \cite{Lutz2}. Because
our model is restricted to s-wave vector-meson nucleon final
states, this expansion takes the simple form,

\begin{eqnarray}
T^{(I)}_{(\pi N\rightarrow V N)\, \mu} (\overline{q},q;w)
\,& = \,
M^{(I,J=\frac {1} {2})}_{\pi N \rightarrow V N}(s)\,
\,{Y}_{(J=\frac {1} {2})\,\mu}(\overline{q},q;w)\nonumber \\
&+ \, M^{(I,J=\frac {3} {2})}_{\pi N \rightarrow V N}(s)\,
\,{Y}_{(J=\frac {3} {2})\,\mu}(\overline{q},q;w),
\label{eq:e10}
\end{eqnarray}

where V stands for $\rho$ or $\omega$ and the angular momentum
projectors are defined by \cite{Lutz1}

\begin{eqnarray}
{Y}_{(J=\frac{1}{2})\,\mu}(\overline{q},q;w)\, = \,
{- \frac {1} {2 \sqrt 3} \,
(\gamma_\mu - \frac {w_\mu} {w^2}\, w \ \mkern-23mu\not \mkern14mu )
(1 - \frac {w \ \mkern-23mu\not} {\sqrt {w^2}})\, i\gamma_5\,},
\label{eq:e11}
\end{eqnarray}

\begin{eqnarray}
{Y}_{(J=\frac {3}{2})\,\mu}&(\overline{q},q;w)\, = \,
- \frac {\sqrt 3} {2} \, (1 + \frac {w \ \mkern-23mu\not} {\sqrt {w^2}})
(q_\mu - \frac {w.q} {w^2} w_\mu ) \, i\gamma_5\nonumber \\
&+ \frac {1} {2 \sqrt 3} \,(\gamma_\mu - \frac {w_\mu} {w^2}\, w \ \mkern-23mu\not
\mkern14mu )\,(1 - \frac {w \ \mkern-23mu\not} {\sqrt {w^2}})
\,(q \ \mkern-23mu\not \mkern10mu- \frac {w.q} {w^2}\, w \ \mkern-23mu\not
\mkern14mu )\, i\gamma_5.
\label{eq:e12}
\end{eqnarray}
In the more general notation of Ref. \cite{Lutz1}, the quantities defined by Eqs. (\ref{eq:e11}) and (\ref{eq:e12})
are written  as $[{Y}_{0,\mu}^{(+)}(\overline{q},q;w)]_{13}$
and $[{Y}_{1,\mu}^{(-)}(\overline{q},q;w)]_{13}$ respectively.

In the center of mass system ($\vec{q}$=$\,-\vec{p}$), the time component of
the spin
projectors ${Y}_{(J=\frac{1}{2})\, \mu}$ and
${Y}_{(J=\frac{3}{2})\, \mu}$ vanishes and the
space components read simply

\begin{eqnarray}
{Y}_{(J=\frac{1}{2})\,j}
\, = \,  \frac {i} {\sqrt 3} \, \pmatrix{\sigma_j
 & 0 \cr 0 & 0\cr},
\label{eq:e13}
\end{eqnarray}

\begin{eqnarray}
{Y}_{(J=\frac{3}{2})\,j}
\, = \,  {i \sqrt 3 \, p_j} \, \pmatrix{0 & 1
\cr 0 & 0\cr}
-\frac {i} {\sqrt3 }\, \pmatrix{ 0 & \sigma_j ({\vec \sigma .\vec p}) \cr 0 & 0\cr}.
\label{eq:e14}
\end{eqnarray}

In these coordinates, the matrix elements of the spin projectors acting
on Dirac spinors can be expressed in terms of Pauli spinors $\chi$ as

\begin{eqnarray}
\overline{u}(\overline{p},\overline{\lambda}) {Y}_{(J=\frac{1}{2})\,j}\, u(p,\lambda)\,= \, i
\frac {\sqrt {p^0 + M_N}
\,\sqrt { \overline{p}^0+
M_N}}  {2  \sqrt 3 M_N} \,\, \overline{\chi} (\overline{\lambda})\, \sigma_j \, \chi (\lambda),
\label{eq:e15}
\end{eqnarray}

\begin{eqnarray}
\overline{u}(\overline{p},\overline{\lambda}) {Y}_{(J=\frac{3}{2})\,j}\, u(p,\lambda)\,= \,
\frac {i\sqrt 3} {2} \,
\frac {\sqrt {\overline{p}^0 + M_N}}
{\sqrt {{p^0} + M_N}}  \,\, \overline{\chi} (\overline{\lambda}) \, \frac {(\vec \sigma . \vec p)} {M_N} \,
\frac {p^j} {m_\pi} \,   \chi (\lambda)\nonumber \\
-\frac {i} {2 \sqrt 3} \,\frac {\sqrt {\overline{p}^0 + M_N}} {\sqrt {p^{0} + M_N}}
 \,\frac { \vec{p}\,^2} {M_N} \,\,\overline{\chi} (\overline{\lambda}) \, \sigma^j \,   \chi (\lambda),
\label{eq:e16}
\end{eqnarray}

where $\lambda$ and $\overline{\lambda}$ are the polarizations of the ingoing and outgoing
nucleons while $p^0$ and $\overline{p}^0$ are defined by $p^0$=$\sqrt {M_N^2 + \overrightarrow{p}^2}$
and $\overline{p}^0$=$\sqrt {M_N^2 + \overrightarrow{\overline{p}}^2}$.
The $\pi N \rightarrow \rho N$ and $\pi N \rightarrow
\omega N$ amplitudes in the S$_{11}$, S$_{31}$, D$_{13}$ and D$_{33}$
channels obtained in Ref. \cite{Lutz1} are  displayed
in Figs. 2 and 3. The quantities shown
are the amplitudes $M^{(I,J)}_{\pi N\rightarrow \rho N}(s)$
and $M^{(I,J)}_{\pi N\rightarrow \omega N}(s)$ defined by
Eq. (\ref {eq:e10}), which depend only on the center of mass
energy $\sqrt s$.
The coupling to subthreshold resonances is clearly
exhi\-bited in these pictures.

In the S$_{11}$ channel, the N(1535) and the N(1650) resonances lead
to peak structures in the imaginary parts of the amplitudes.
The pion-induced $\omega$ production amplitudes in the D$_{13}$ channel
reflect the strong coupling of the N(1520) re\-sonance to the
$\omega$N channel. The $\pi N \rightarrow \omega N$ amplitudes
contain also significant contributions from non-resonant, background
terms. We refer to \cite{Lutz1} for a detailed discussion of the
dynamical structure of the amplitudes and for the comparison to
quark-model coupling constants of
the effective coupling strengths of the vector meson-nucleon channels
to the baryon resonances contributing to the amplitudes.
\newpage

\begin{figure}[t]
\noindent
\begin{center}
\mbox{\epsfig{file=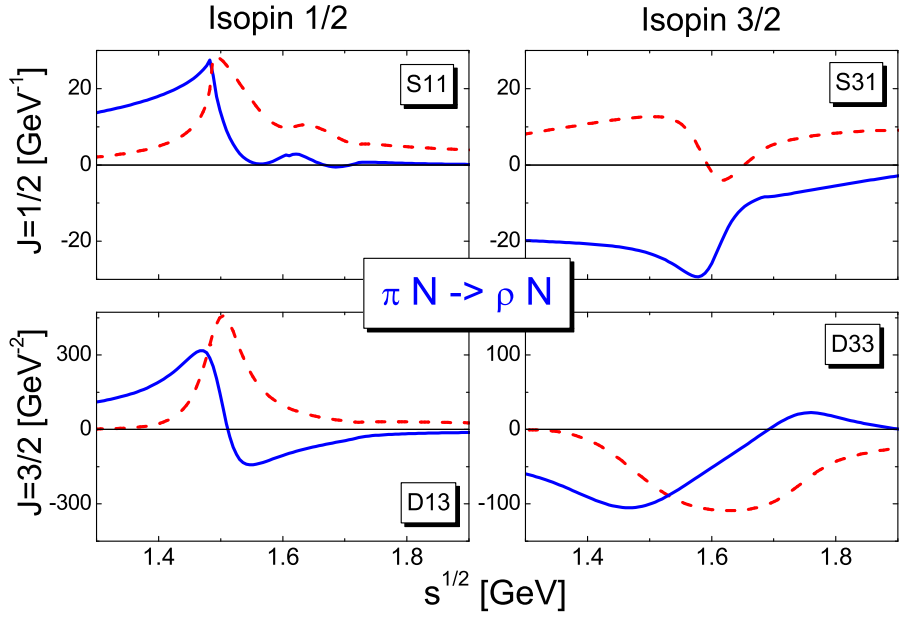, height=9.5 cm}}
\end{center}
\caption{Real and imaginary parts of
the $\pi N \rightarrow \rho^0 N$ amplitudes
in the pion-nucleon
S$_{11}$, S$_{31}$, D$_{13}$ and D$_{33}$
partial waves [16].}
\label{f2}
\end{figure}
\vglue 0.05 true cm
\begin{figure}[h]
\begin{center}
\mbox{\epsfig{file=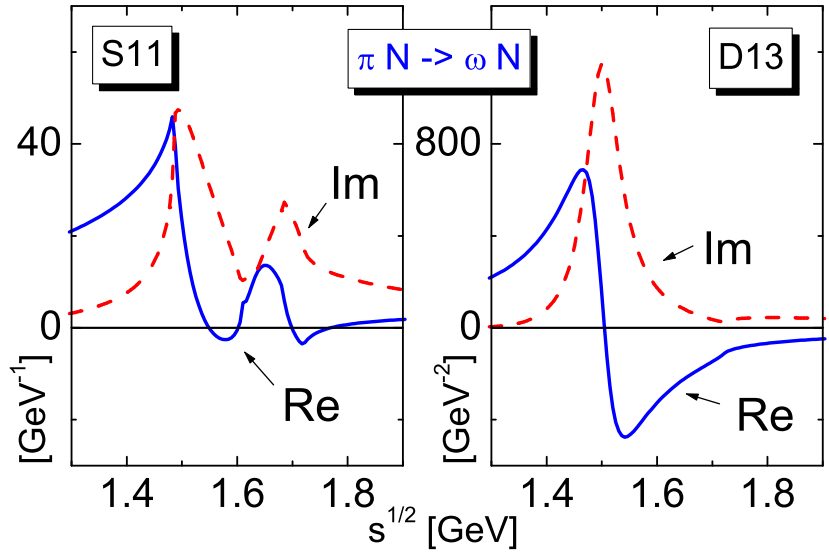, height=8.5 cm}}
\end{center}
\caption{Real and imaginary parts of
the $\pi N \rightarrow \omega N$ amplitudes in the pion-nucleon
S$_{11}$ and D$_{13}$ partial waves [16].}
\label{f3}
\end{figure}

\newpage

The $\pi^-p \rightarrow \rho^0 n$ and $\pi^-p \rightarrow \omega n$
amplitudes are obtained from the
isospin 1/2 and isospin 3/2 scattering amplitudes by the
relations,
\noindent
\begin{eqnarray}
M^J_{\pi^-p \rightarrow \rho^0 n} = - \frac {\sqrt 2} {3}
M^{(1/2,J)}_{\pi N\rightarrow \rho N}  +\frac {\sqrt 2} {3}
M^{(3/2,J)}_{\pi N\rightarrow \rho N},
\label{eq:e17}
\end{eqnarray}
\noindent
\begin{eqnarray}
M^J_{\pi^-p \rightarrow \omega n} = \sqrt{\frac {2} {3}}
M^{(1/2,J)}_{\pi N\rightarrow \omega N}.
\label{eq:e18}
\end{eqnarray}

\noindent
Similarly the $\pi^+ n \rightarrow \rho^0 p$ and $\pi^+n \rightarrow \omega p$
amplitudes are given by
\noindent
\begin{eqnarray}
M^J_{\pi^+n \rightarrow \rho^0 p} = \frac {\sqrt 2} {3}
M^{(1/2,J)}_{\pi N\rightarrow \rho N}  -\frac {\sqrt 2} {3}
M^{(3/2,J)}_{\pi N\rightarrow \rho N},
\label{eq:e19}
\end{eqnarray}
\noindent
\begin{eqnarray}
M^J_{\pi^+n \rightarrow \omega p} = \sqrt{\frac {2} {3}}
M^{(1/2,J)}_{\pi N\rightarrow \omega N}.
\label{eq:e20}
\end{eqnarray}

\noindent
The quantities $M^J_{\pi^-p \rightarrow \rho^0 n}$
and $M^J_{\pi^-p \rightarrow \omega n}$ are displayed in Fig. 4.
Their counterparts for the other reactions,
$M^J_{\pi^+n \rightarrow \rho^0 p}$
and $M^J_{\pi^+n \rightarrow \omega p}$~, are shown in Fig.~5.

\noindent
The phases of the isospin coefficients appearing in Eqs. (\ref {eq:e17}) and (\ref {eq:e19}) play a crucial
role in determining the $\rho^0-\omega$ interference in the
$\pi^-p \rightarrow e^+e^- n$ and $\pi^+n \rightarrow e^+e^- p$
reaction cross sections.
The real and imaginary parts of the $\pi^-p \rightarrow \omega n$
and of the $\pi^+n \rightarrow \omega p$ amplitudes are the same and mostly positive.
In contrast, the $\pi^-p \rightarrow \rho^0 n$ and $\pi^+n \rightarrow \rho^0 p$ amplitudes
have opposite signs. The $\pi^-p \rightarrow \rho^0 n$ amplitudes are predominantly
negative and will therefore interfere destructively with the $\pi^-p \rightarrow \omega n$
amplitudes. The $\pi^+n \rightarrow  \rho^0 p$ and $\pi^+n \rightarrow \omega p$
amplitudes have the same sign over a large $\sqrt s$ interval, leading to a constructive
interference.

From Figs. 4 and 5, we expect the $\pi^-p \rightarrow e^+e^- n$ and
$\pi^+n \rightarrow e^+e^- p$ reaction cross sections to be very sensitive
to the presence of baryon resonances below the vector meson production threshold.

\newpage
\begin{figure}[t]
\begin{center}
\mbox{\epsfig{file=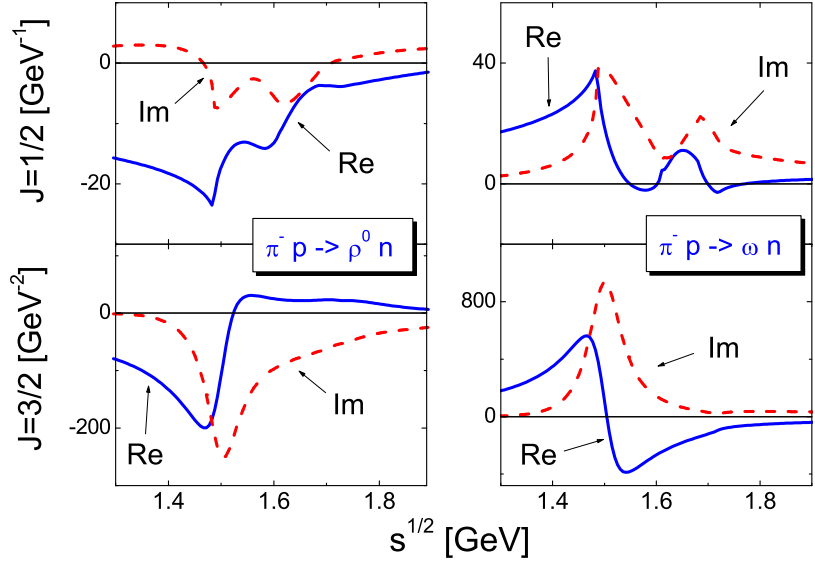, height= 8.7cm}}
\end{center}
\caption{Scattering amplitudes $M^J_{\pi^-p \rightarrow \rho^0 n}$
and $M^J_{\pi^-p \rightarrow \omega n}$
obtained from the model of Ref. [16]
for the spin J=1/2 and J=3/2 channels.}
\vskip 0.4truecm
\label{f4}
\end{figure}
\begin{figure}[h]
\begin{center}
\mbox{\epsfig{file=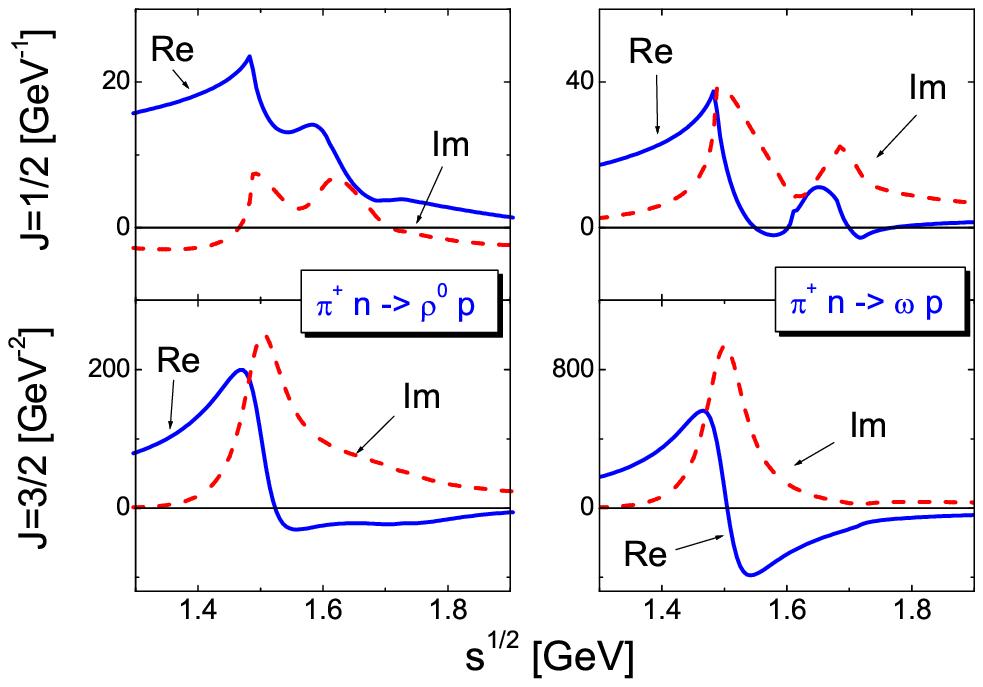, height= 8.7cm}}
\end{center}
\caption{Scattering amplitudes $M^J_{\pi^+n \rightarrow \rho^0 p}$
and $M^J_{\pi^+n \rightarrow \omega p}$
obtained from the model of Ref. [16]
for the spin J=1/2 and J=3/2 channels.}
\vskip 0.4truecm
\label{f5}
\end{figure}
\par
\newpage

\section{Calculation of the $\pi^-p \rightarrow e^+e^- n$ and $\pi^+n \rightarrow e^+e^- p$ cross sections
close to the vector
meson production threshold}

The $\pi^-p \rightarrow e^+e^- n$ and $\pi^+n \rightarrow e^+e^- p$ cross sections are calculated from
the $\pi^-p \rightarrow \rho^0 n$, $\pi^-p \rightarrow \omega n$,
$\pi^+n \rightarrow \rho^0 p$ and $\pi^+n \rightarrow \omega p$
amplitudes presented in Section 2, supplemented with the assumption of
the Vector Meson Dominance of the electromagnetic current \cite{Sakurai,Kroll}.
This assumption can be enforced in the effective Lagrangian by introducing
vector meson-photon interaction terms of the form,
\begin{eqnarray}
{\mathcal L^{int}_{\gamma V}}\,&=&\, \frac {f_\rho} {2 M_\rho^2} F^{\mu \nu}\, \rho^0_{\mu \nu}
\,+\,\frac {f_\omega} {2 M_\omega^2} F^{\mu \nu}\, \omega_{\mu \nu},
\label{eq:e21}
\end{eqnarray}
\noindent
where the photon and vector meson field tensors are defined by
\begin{eqnarray}
F^{\mu \nu}\, =\,\partial^\mu A^\nu -\partial^\nu A^\mu,
\label{eq:e22}
\end{eqnarray}
\begin{eqnarray}
V^{\mu \nu}\, =\,\partial^\mu V^\nu -\partial^\nu V^\mu.
\label{eq:e23}
\end{eqnarray}
In equation (\ref {eq:e21}),
$M_\rho$ and $M_\omega$ are the $\rho$- and $\omega$-masses and
$f_\rho$ and  $f_\omega$ are dimensional coupling constants. Their magnitude can be
determined from the $e^+e^-$ partial decay widths of the $\rho$- and
$\omega$-mesons to be \cite{Friman1}
\begin{eqnarray}
|f_\rho|= 0.036\, GeV^2,
\label{eq:e24}
\end{eqnarray}
\begin{eqnarray}
|f_\omega|= 0.011 \,GeV^2.
\label{eq:e25}
\end{eqnarray}
The relative sign of $f_\rho$ and $f_\omega$
is fixed by vector meson photoproduction amplitudes \cite{Lutz1}.
We assume that the phase correlation between isoscalar and isovector currents
is identical for real and virtual photons as in Sakurai's realization
of the Vector Meson Dominance assumption \cite{Sakurai}. With the conventions used
in this paper, both $f_\rho$ and $f_\omega$ are positive.
The form of the coupling terms of Eq. (\ref {eq:e21}) is
appropriate for describing the hadronic structure of massive photons.
%These terms vanish for real photons which couple to vector fields through different gauge
%invariant interactions \cite{Lutz1}.

We consider first the $\pi^-p \rightarrow  e^+e^- n$ reaction. The diagrams contributing to
this process in the Vector Meson
Do\-minance model are shown in Fig. 6.

\begin{figure}[h]
\begin{center}
\mbox{\epsfig{file=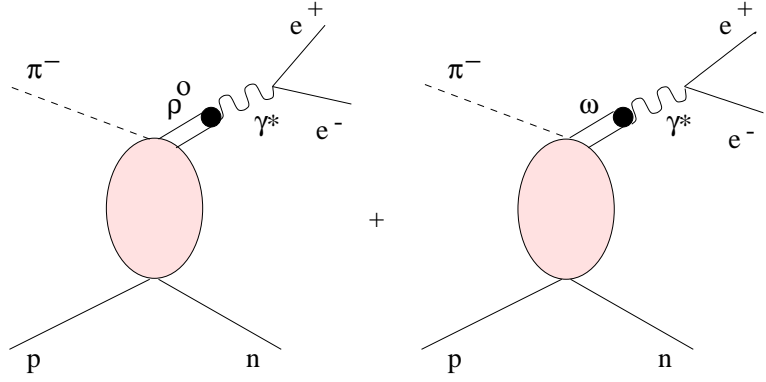, height=5cm}}
\end{center}
\caption{Diagrams contributing to the $\pi^-p \rightarrow  e^+e^- n$
amplitude with intermediate $\rho^0$- and $\omega$-mesons.}
\vskip 0.4truecm
\label{f6}
\end{figure}

We denote the momenta of the ingoing and outgoing hadrons as in Fig. \ref{f1}
and the 4-momenta of the electron and the positron
by $p_-=(p^0_-, \vec p_-)$ and $p_+=(p^0_+, \vec p_+)$ respectively.
The diffe\-rential cross section for the $\pi^-p \rightarrow e^+e^- n$ reaction
in the center of mass reference frame is then given by

\begin{eqnarray}
 \biggl[\frac{d\sigma}{d\overline{q}^2}\biggr]_{\pi^- p \rightarrow e^+e^- n} =
\, \frac {M_p\,M_n} {16 \pi^2 s}\, \frac {|\vec {\overline{p}}|} {|\vec p|}
\int \frac {d^3 \vec p_+}{(2\pi)^3}\, \frac {m_e}{p_+^0}\,
\int \frac {d^3 \vec p_-}{(2\pi)^3}\, \frac {m_e}{p_-^0} \,(2\pi)^4 \,
& \nonumber \\
\sum_{\lambda,\overline{\lambda},\lambda_+,\lambda_-} |\mathcal M
_{\pi^- p \rightarrow e^+e^- n }
(q,p,\lambda;p_+,\lambda_+,p_-,\lambda_-,\overline{p},\overline{\lambda})|\,^2
\, \delta^4 (\overline{q}-p_+-&p_-)  ,
\label{eq:e26}
\end{eqnarray}
\noindent
in which $m_e$ denotes the electron mass and where the magnitude of the initial and final nucleon momenta
is given as function of $\sqrt s$ and $\overline{q}^2$ by
\begin{eqnarray}
|\vec {p}|\,=\, \frac {\sqrt s} {2} \, \Bigl[ 1 -2\  \frac {M_p^2 + m_\pi^2}{s}
+\frac {(M_p^2 - m_\pi^2)^2}{s^2} \, \Bigr]^{\frac {1} {2}},
\label{eq:e32}
\end{eqnarray}
\noindent
\begin{eqnarray}
|\vec {\overline{p}}|\,=\, \frac {\sqrt s} {2} \, \Bigl[ 1 -2\  \frac {M_p^2 + \overline{q}^2}{s}
+\frac {(M_p^2 - \overline{q}^2)^2}{s^2} \, \Bigr]^{\frac {1} {2}}.
\label{eq:e33}
\end{eqnarray} .

We factorize the invariant matrix elements into vector meson
production and $e^+e^-$ decay amplitudes as
\begin{eqnarray}
\mathcal M  _{\pi^- p \rightarrow e^+e^- n }
(q,p,\lambda;p_+,\lambda_+,p_-,\lambda_-,\overline{p},\overline{\lambda})
\,=
\mkern -50 mu &\nonumber \\
\mathcal M ^\mu_{\pi^- p \rightarrow \rho^0 n } (q,p,\lambda;\overline{q},\overline{p},\overline{\lambda})\,
\mathcal M & _{\rho^0 \rightarrow e^+e^-\ \mu } \,(\overline{q};p_+,\lambda_+,p_-,\lambda_-)\nonumber \\
+\,\mathcal M ^\mu_{\pi^- p \rightarrow \omega n }(q,p,\lambda;\overline{q},\overline{p},\overline{\lambda})\,
\mathcal M & _{\omega \rightarrow e^+e^- \ \mu} \,(\overline{q};p_+,\lambda_+,p_-,\lambda_-).
\label{eq:e27}
\end{eqnarray}
\newpage
\noindent
The $e^+e^-$ decay amplitudes include the vector meson propagators,
\begin{eqnarray}
S_\rho(\overline{q}^2)\, \equiv \, \frac {1} {\overline{q}^2-M_\rho^2+i\Gamma_\rho(\overline{q}^2)\,M_\rho},
\label{eq:e28}
\end{eqnarray}
\begin{eqnarray}
S_\omega(\overline{q}^2) \,\equiv \, \frac {1} {\overline{q}^2-M_\omega^2+i\Gamma_\omega \,M_\omega},
\label{eq:e29}
\end{eqnarray}
\noindent
where the energy-dependent $\rho$-width is given by
\begin{eqnarray}
\Gamma_\rho(\overline{q}^2)\, =\,\Gamma_\rho\, \, \frac {M_\rho} {\sqrt{\overline{q}^2}}
\, \biggl( \frac {\overline{q}^2 - 4 m_\pi^2} {M_\rho^2 - 4 m_\pi^2}\biggr)^{\frac {3} {2}},
\label{eq:e30}
\end{eqnarray}
\noindent
$\Gamma_\rho$ and $\Gamma_\omega$ denoting the widths at the
peak of the $\rho$- and $\omega$-resonances.

We perform the lepton sums and integrations and average over the
angle between the initial and final 3-momenta. Using Eqs. (\ref {eq:e11}) to (\ref {eq:e16}), we find

\begin{eqnarray} \displaystyle
\biggl[\frac{d\sigma}{d\overline{q}^2}\biggr]_{\pi^- p \rightarrow e^+e^- n} =
\,\frac {\alpha} {6 \pi^2}\,\frac {M_p\,M_n} {s}\,\frac {|\vec {\overline{p}}|} {|\vec
p|} \,  {m_e^2} (1+\frac {\overline{q}^2} {2m_e^2} )\,
(1-\frac {4m_e^2}{\overline{q}^2}  )^{\frac {1} {2}}&
\nonumber \\
\Biggl[\,\frac{ f_\rho^2} {M_\rho^4} \ S^*_\rho(\overline{q}^2) \,S_\rho(\overline{q}^2)\
\sum_{J} \overline{C}_{JJ}\
 M ^{J*}_{\pi^- p \rightarrow \rho^0 n } (s)\,
 M ^J_{\pi^- p \rightarrow \rho^0 n} (s)\,
& \nonumber \\
+\, \frac {f_\rho f_\omega}{M_\rho^2 M_\omega^2}  \ S^*_\rho(\overline{q}^2) \,S_\omega(\overline{q}^2)\
\sum_{J} \overline{C}_{JJ}\
M ^{J*}_{\pi^- p \rightarrow \rho^0 n } (s)\,
M ^J_{\pi^- p \rightarrow \omega n } (s)\,
& \nonumber \\
+\, \frac {f_\omega f_\rho} {M_\rho^2 M_\omega^2} \ S^*_\omega(\overline{q}^2) \,S_\rho(\overline{q}^2)\
 \sum_{J} \overline{C}_{JJ}\
M ^{J*}_{\pi^- p \rightarrow \omega n } (s)\,
M ^J_{\pi^- p \rightarrow \rho^0 n } (s)\,
& \nonumber \\
+\, \frac {f_\omega^2} {M_\omega^4} \, S^*_\omega(\overline{q}^2) \,S_\omega(\overline{q}^2)\,
 \sum_{J} \overline{C}_{JJ}\
M ^{J*}_{\pi^- p \rightarrow \omega^0 n } (s)\,
M^J _{\pi^- p \rightarrow \omega n } (s)\,\Biggr], &
\label{eq:e34}
\end{eqnarray}
\noindent
where the $\overline{C}_{JJ}$ coefficients depend on $\sqrt s$ and $\overline{q}^2$
and are given
in the center of
mass frame by the expressions,

\noindent
\begin{eqnarray}
\overline{C}_{\frac {1}{2}\, \frac {1}{2}}\ = \frac {(p^0 + M_p)
\,(\overline{p}^0+M_n)}  { 4 M_nM_p}
\,\Bigl( 1+\frac {|\vec {\overline{p}}|^2} {3 \overline{q}^2}\Bigr),
\label{eq:e35}
\end{eqnarray}
\noindent
\begin{eqnarray}
\overline{C}_{\frac {3}{2}\, \frac {3}{2}}\ =
\frac {(\overline{p}^0+M_n)} {(p^0 + M_p)}
\, \frac {|\vec p|^4} {2 M_n M_p}
\,\Bigl( 1+\frac {|\vec {\overline{p}}|^2} {3 \overline{q}^2}\Bigr).
\label{eq:e36}
\end{eqnarray}

\noindent

The derivation of the cross section in the other isospin channel, $\pi^+n \rightarrow e^+e^- p$,
is completely similar, with the obvious replacement of
$\mathcal M _{\pi^- p \rightarrow \rho^0 n  \ \mu}$ and
$\mathcal M _{\pi^- p \rightarrow \omega n  \ \mu}$
by $\mathcal M _{\pi^+ n \rightarrow \rho^0 p  \ \mu}$ and
$\mathcal M _{\pi^+ n \rightarrow \omega p  \ \mu}$.

\section{Numerical results}

This section is devoted to the discussion of the
$\pi^-p \rightarrow e^+e^- n$ and $\pi^+n \rightarrow e^+e^- p$ differential cross sections
for values of the total center of mass energy $\sqrt s$
ranging from 1.5 GeV up to 1.8 GeV. We explore the dependence of the $\rho^0-\omega$ interference
pattern in the $e^+e^-$ channel on $\sqrt s$ in this energy range, in particular
in the vicinity of the $\omega$-meson production threshold ($\sqrt s$=1.72 GeV).

We consider first the differential cross section defined by Eq. (\ref {eq:e34})
for the $\pi^-p \rightarrow e^+e^- n$
and the $\pi^+n \rightarrow e^+e^- p$ reactions. The magnitude of the 4-vector $\overline{q}$
is the invariant mass m$_{e^+e^-}$ of the $e^+e^-$ pair.

The differential cross sections calculated using Eq. (\ref{eq:e34}) for the $\pi^-p \rightarrow e^+e^- n$
and the $\pi^+n \rightarrow e^+e^- p$ reactions at $\sqrt s$=1.5 GeV
are shown in Figs. 7 and ~8.

These figures illustrate very clearly
the isospin effects discussed in Section 2.
For the two reactions, the $\omega$ and $\rho^0$ contributions to the cross section are the
same.
The $\rho^0$-$\omega$ interference is destructive for the $\pi^-p \rightarrow e^+e^- n$
reaction and constructive for the $\pi^+n \rightarrow e^+e^- p$ process.
Consequently, the $\pi^-p \rightarrow e^+e^- n$ differential cross section is extremely small
in the range of invariant masses considered in this calculation (less than 10 nb GeV$^{-2}$).
In contrast, the constructive $\rho^0$-$\omega$ interference for the $\pi^+n \rightarrow e^+e^- p$
reaction leads to a sizeable differential cross section (of the order of 0.15 $\mu$b GeV$^{-2}$).

This is a very striking prediction, linked to the resonant structure of the scattering
amplitudes $M^{1/2}_{\pi N\rightarrow V N}$ and
$M^{3/2}_{\pi N\rightarrow V N}$. At $\sqrt s$=1.5 GeV, the coefficients
$\overline{C}_{\frac {1}{2}\, \frac {1}{2}}$ and $\overline{C}_{\frac {3}{2}\, \frac {3}{2}}$
are 1.12 and 0.02 GeV$^2$ respectively (for $e^+e^-$ pair invariant masses of the order
of 0.5 GeV). The smallness of $\overline{C}_{\frac {3}{2}\, \frac {3}{2}}$
is a consequence of the relative D-wave state in the initial pion-nucleon system
implied by the spin 3/2 of the channel.
From the values of the amplitudes displayed in Figs. 4 and 5, it is easy
to see that the J=1/2 and J=3/2 contributions to the $\pi^-p \rightarrow e^+e^- n$
and $\pi^+n \rightarrow e^+e^- p$ differential cross sections are of comparable magnitude.
These cross sections reflect the couplings of both the N(1520) and N(1535) baryon
resonances to the
vector meson-nucleon channels. Data on differential cross sections for the $\pi^-p \rightarrow e^+e^- n$
and $\pi^+n \rightarrow e^+e^- p$ reactions at $\sqrt s$=1.5 GeV would be very
useful for making progress in the understanding of these couplings.
\newpage

\begin{figure}[t]
\begin{center}
\mbox{\epsfig{file=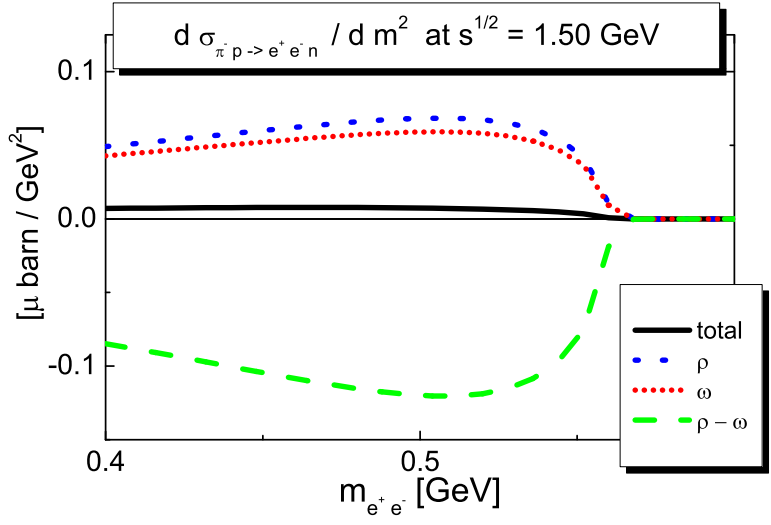, height= 8cm}}
\end{center}
\caption{Differential cross section for the $\pi^-p \rightarrow e^+e^- n$
reaction at $\sqrt s$=1.5 GeV as function of the invariant mass of the $e^+e^-$
pair. The $\rho^0$ and the $\omega$ contributions are indicated by short-dashed
and dotted lines respectively. The long-dashed line shows the $\rho^0-\omega$ interference.
The solid line is the sum of the three contributions. }
\vskip 0.2truecm
\label{f7}
\end{figure}

\begin{figure}[h]
\begin{center}
\mbox{\epsfig{file=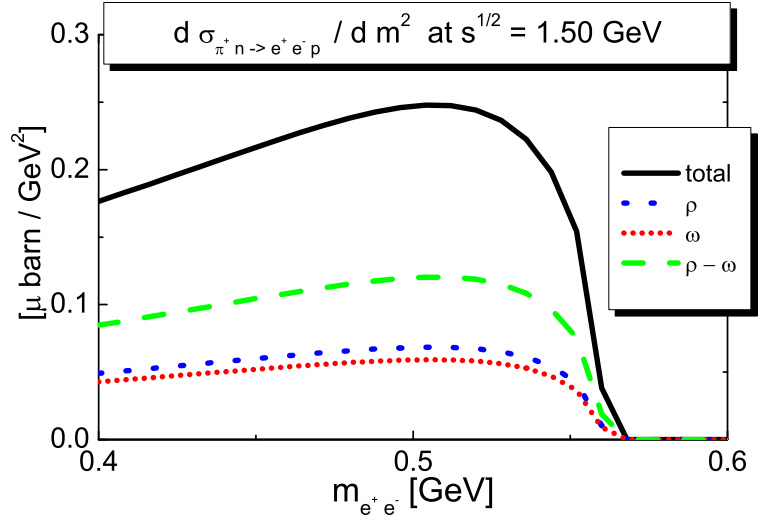, height= 8cm}}
\end{center}
\caption{Differential cross section for the $\pi^+n \rightarrow e^+e^- p$
reaction at $\sqrt s$=1.5 GeV as function of the invariant mass of the $e^+e^-$
pair. The $\rho^0$ and the $\omega$ contributions are indicated by short-dashed
and dotted lines respectively. The long-dashed line shows the $\rho^0-\omega$ interference.
The solid line is the sum of the three contributions.}
\label{f8}
\end{figure}
\par

\newpage
The $\sqrt s$-dependence of the $\pi^-p \rightarrow e^+e^- n$ differential cross section
below the vector meson production threshold is illustrated in Figs. 9-11. The correspon\-ding results for
the $\pi^+n \rightarrow e^+e^- p$ reaction are shown in Figs. 12-14.

\begin{figure}[h]
\begin{center}
\mbox{\epsfig{file=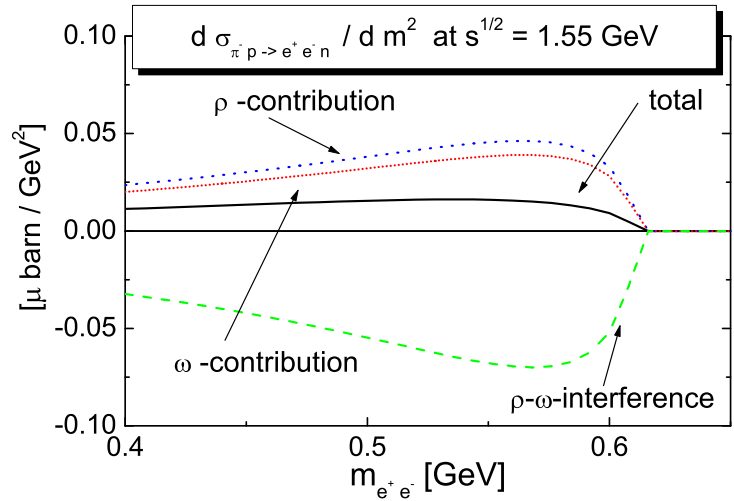, height= 7.7cm}}
\end{center}
\caption{Differential cross section for the $\pi^-p \rightarrow e^+e^- n$
reaction at $\sqrt s$=1.55 GeV as function of the invariant mass of the $e^+e^-$
pair.}
\label{f9}
\end{figure}
\smallskip
\begin{figure}[h]
\begin{center}
\mbox{\epsfig{file=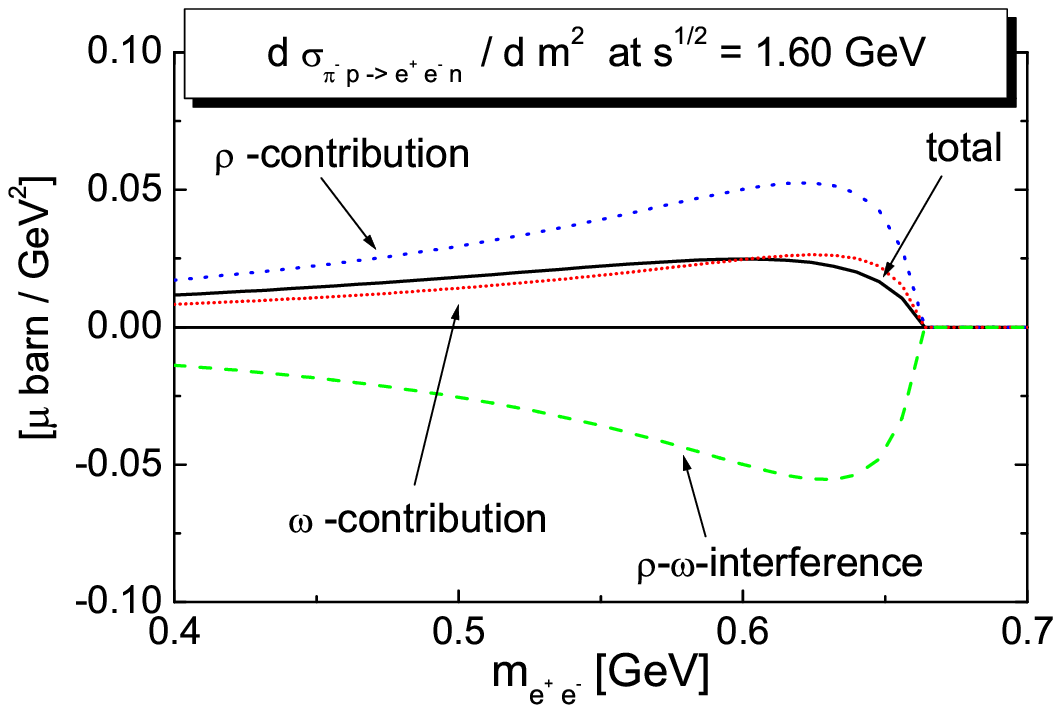, height= 7.7cm}}
\end{center}
\caption{Differential cross section for the $\pi^-p \rightarrow e^+e^- n$
reaction at $\sqrt s$=1.60 GeV as function of the invariant mass of the $e^+e^-$
pair.}
\label{f10}
\end{figure}
\newpage

\smallskip
\begin{figure}[t]
\begin{center}
\mbox{\epsfig{file=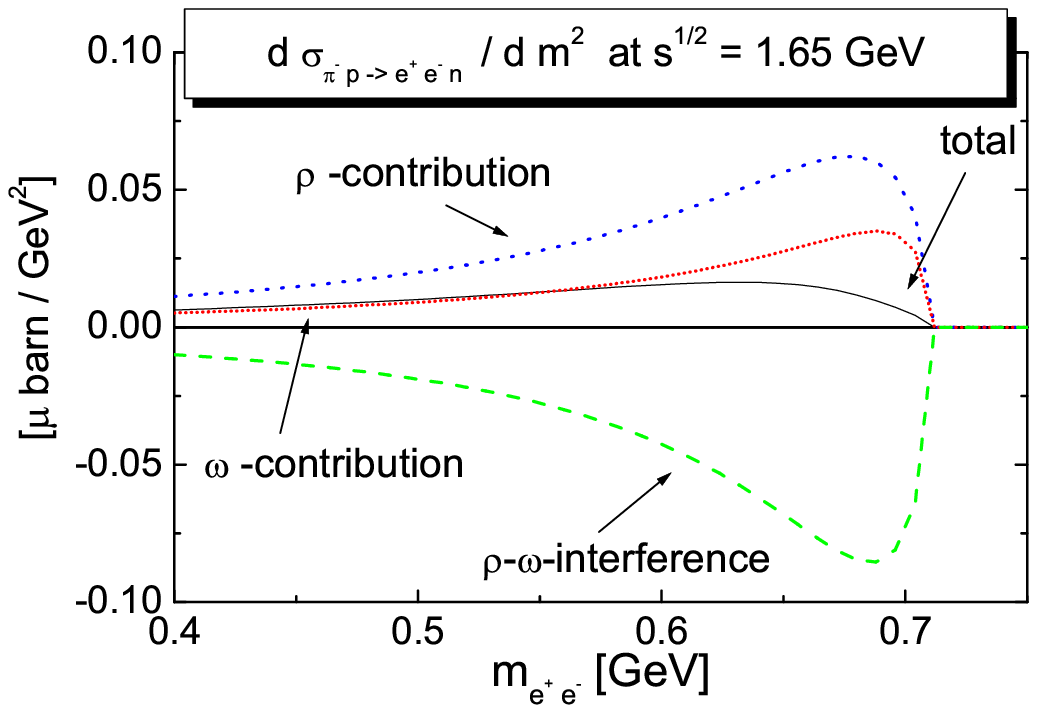, height= 7.7cm}}
\end{center}
\caption{Differential cross section for the $\pi^-p \rightarrow e^+e^- n$
reaction at $\sqrt s$=1.65 GeV as function of the invariant mass of the $e^+e^-$
pair.}
\label{f11}
\end{figure}

\begin{figure}[t]
\begin{center}
\mbox{\epsfig{file=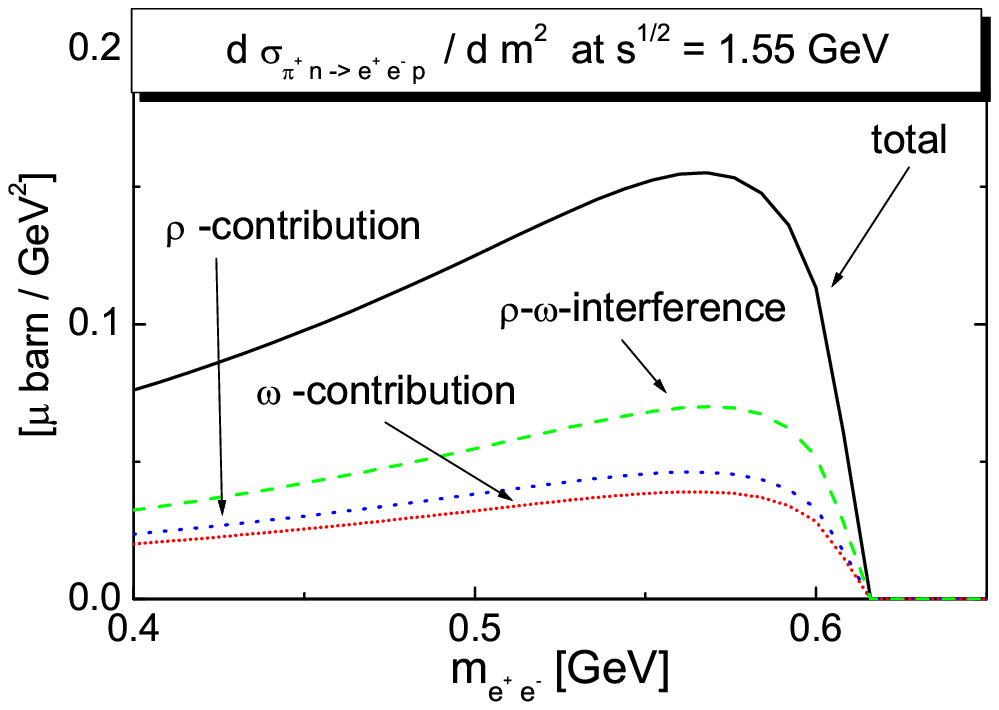, height= 7.7cm}}
\end{center}
\caption{Differential cross section for the $\pi^+n \rightarrow e^+e^- p$
reaction at $\sqrt s$=1.55 GeV as function of the invariant mass of the $e^+e^-$
pair.}
\label{f12}
\end{figure}
\smallskip
\begin{figure}[h]
\begin{center}
\mbox{\epsfig{file=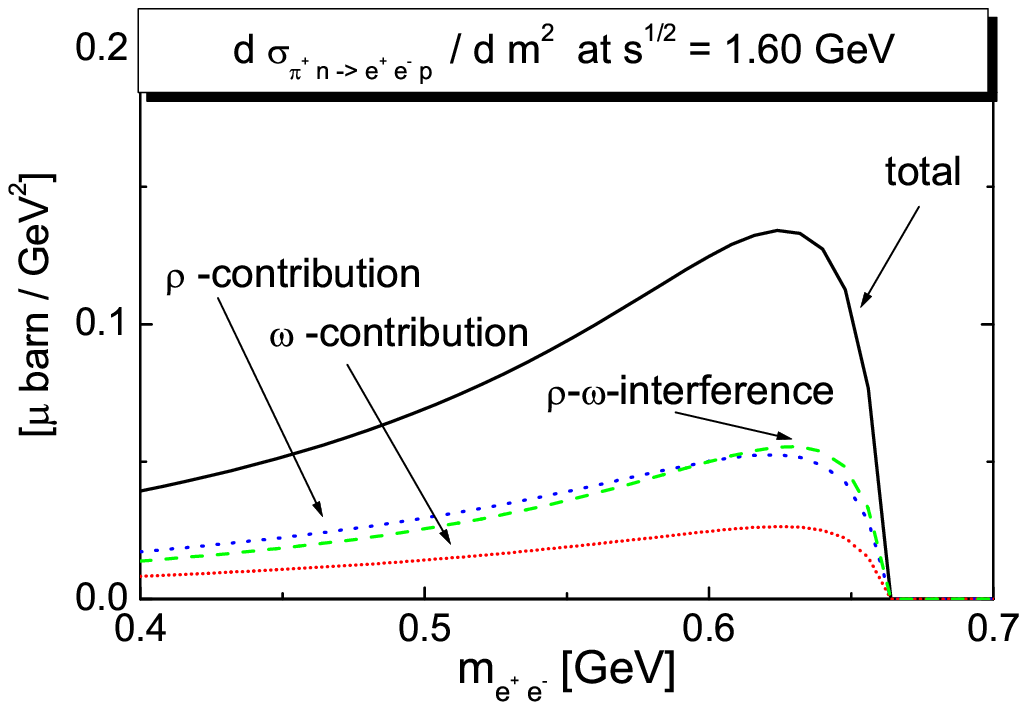, height= 7.7cm}}
\end{center}
\caption{Differential cross section for the $\pi^+n \rightarrow e^+e^- p$
reaction at $\sqrt s$=1.60 GeV as function of the invariant mass of the $e^+e^-$
pair.}
\label{f13}
\end{figure}
\smallskip
\begin{figure}[h]
\begin{center}
\mbox{\epsfig{file=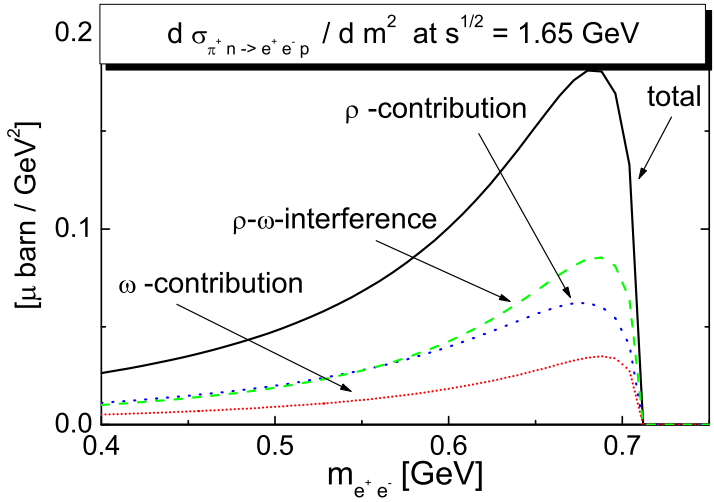, height= 7.7cm}}
\end{center}
\caption{Differential cross section for the $\pi^+n \rightarrow e^+e^- p$
reaction at $\sqrt s$=1.65 GeV as function of the invariant mass of the $e^+e^-$
pair.}
\label{f14}
\end{figure}
\par

These differential
cross sections vary smoothly with the total center of mass energy. They
exhibit the features discussed for $\sqrt s$=1.5 GeV. However they also reflect
dynamics associated with higher-lying resonances \cite{Lutz1}. We emphasize again that
the relative s-wave assumed between the vector meson and the nucleon in the final state
is appropriate only for values of m$_{e^+e^-}$ close to ($\sqrt s - M_N$). The diffe\-rential
cross sections for the lowest $e^+e^-$ pair invariant masses are expected to be outside
the range of validity of the model of Ref. \cite{Lutz1}.

The interference pattern changes drastically at the $\omega$-meson threshold.
This is shown in Figs. 15 and 16 for the $\pi^-p \rightarrow e^+e^- n$ differential cross section
and in Figs. 17 and 18 for the $\pi^+n \rightarrow e^+e^- p$ differential cross section.
\par

\begin{figure}[t]
\begin{center}
\mbox{\epsfig{file=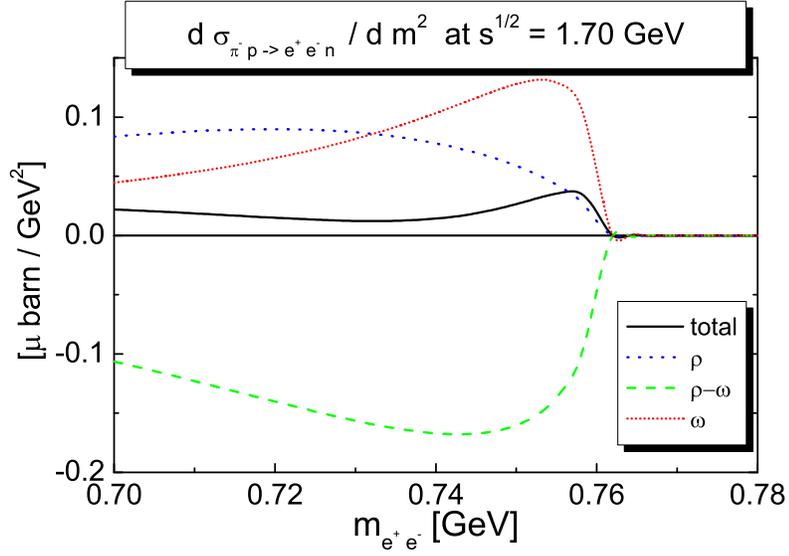, height= 8.5cm}}
\end{center}
\caption{Differential cross section for the $\pi^-p \rightarrow e^+e^- n$
reaction at $\sqrt s$=1.70 GeV as function of the invariant mass of the $e^+e^-$
pair. }
\vskip 0.2truecm
\label{f15}
\end{figure}
\begin{figure}[h]
\begin{center}
\mbox{\epsfig{file=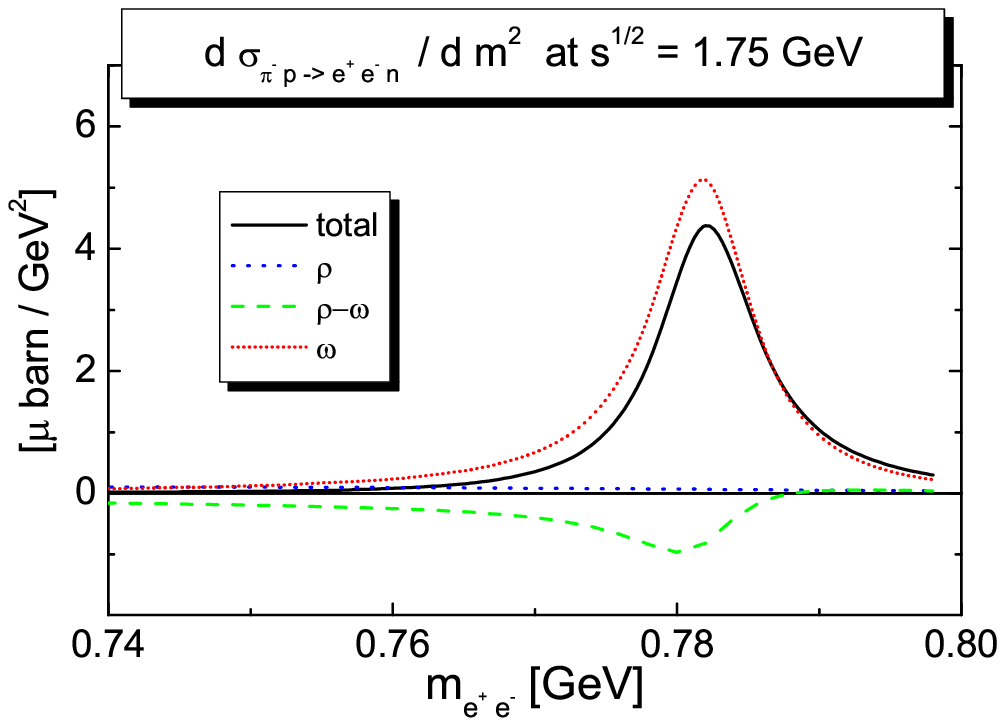, height= 8.5cm}}
\end{center}
\caption{Differential cross section for the $\pi^-p \rightarrow e^+e^- n$
reaction at $\sqrt s$=1.75 GeV as function of the invariant mass of the $e^+e^-$
pair.}
\label{f16}
\end{figure}
\par
\newpage
\begin{figure}[t]
\begin{center}
\mbox{\epsfig{file=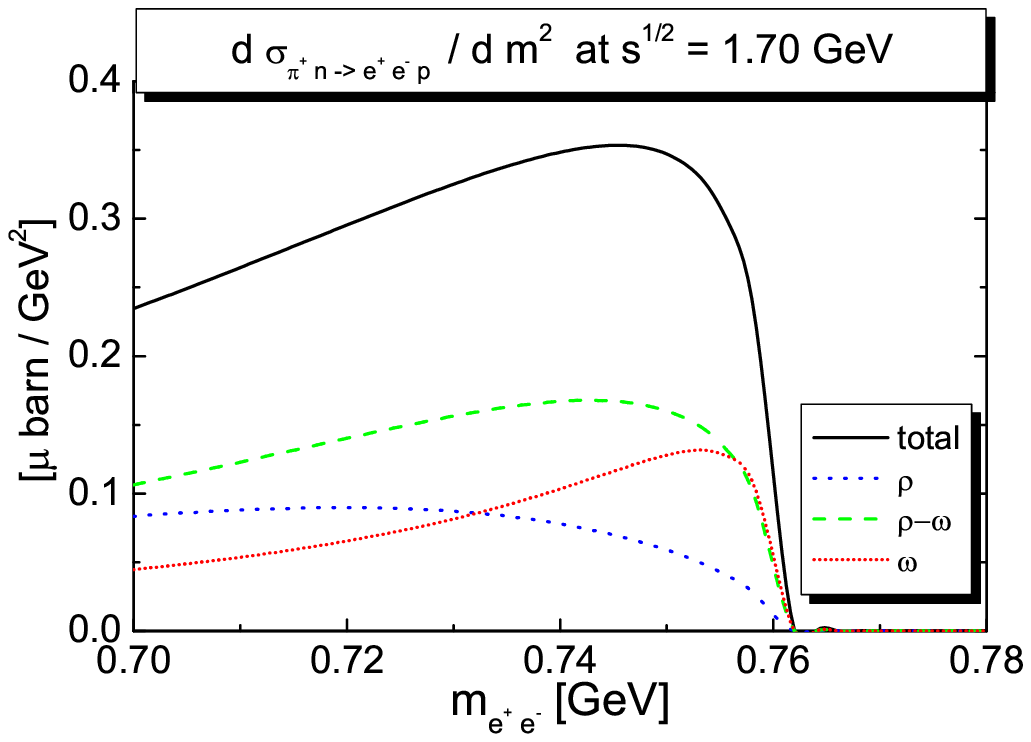, height= 8.5cm}}
\end{center}
\caption{Differential cross section for the $\pi^+n \rightarrow e^+e^- p$
reaction at $\sqrt s$=1.70 GeV as function of the invariant mass of the $e^+e^-$
pair. }
\vskip 0.2truecm
\label{f17}
\end{figure}
\begin{figure}[h]
\begin{center}
\mbox{\epsfig{file=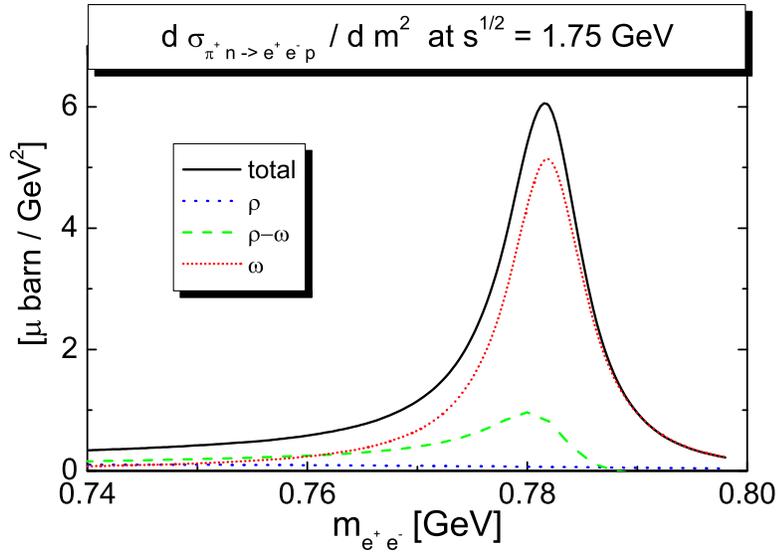, height= 8.5cm}}
\end{center}
\caption{Differential cross section for the $\pi^+n\rightarrow e^+e^- p$
reaction at $\sqrt s$=1.75 GeV as function of the invariant mass of the $e^+e^-$
pair.}
\label{f18}
\end{figure}
\par

Just below threshold, the $\omega$-contribution begins to increase, while the general
features of the $e^+e^-$ production in the two isospin channels remain the same.
Above the $\omega$-meson production threshold,
the differential cross sections for the $\pi^-p \rightarrow e^+e^- n$ and
$\pi^+n\rightarrow e^+e^- p$ reactions are completely dominated by the
$\omega$-contribution. The magnitudes of the cross sections for the two reactions are
comparable. The $\rho^0-\omega$ interference is still destructive
in the $\pi^-p \rightarrow e^+e^- n$ channel and constructive in
the $\pi^+n\rightarrow e^+e^- p$ channel, albeit very small.
In both reactions, crossing the $\omega$-production threshold
leads to a sharp increase in the cross section, by two orders of magnitude
in the $\pi^-p \rightarrow e^+e^- n$ channel and by one order of
magnitude in the $\pi^+n\rightarrow e^+e^- p$ channel.

In order to see the s-channel resonant structure of the $e^+e^-$ pair
production more directly, it is interesting to look at the differential cross section
 as function of $\sqrt s$ for $e^+e^-$ pairs of given invariant mass.
This quantity is displayed in Fig. 19 for the $\pi^-p \rightarrow e^+e^- n$ reaction
and in Fig. 20 for the $\pi^+n\rightarrow e^+e^- p$ reaction.
The invariant mass of the $e^+e^-$ pair is 0.55 GeV. The structures associated with
the N(1520) and N(1535) baryon
resonances are particularly visible. In Figs. 21 and 22, we show
differential cross sections as functions of $\sqrt s$ for $e^+e^-$ pairs
of invariant masses ranging from m$_{e^+e^-}$=0.40 GeV to 0.65 GeV.
Our results are presented in Figs. 21 and 22 for the $\pi^-p \rightarrow e^+e^- n$ and
$\pi^+n\rightarrow e^+e^- p$ reactions respectively.
The N(1650), the $\Delta$(1620) and the $\Delta$(1700) play
a significant role in determining the cross sections for
values of $\sqrt s$ close to the $\omega$-meson production threshold.

It is interesting to compare our cross sections for the $\pi^-p \rightarrow e^+e^- n$
and $\pi^+n\rightarrow e^+e^- p$ reactions with the recent work of Titov
and K\"{a}mpfer \cite{Titov}. In their approach, the $\pi N \rightarrow e^+e^- N$
amplitudes are assumed to be dominated by
s- and u-channel nucleon and baryon resonance exchanges. All baryon re\-sonances
with masses $\leq$1.72 GeV are included. The transition couplings of baryon
resonances to vector fields are taken from the chiral
quark model calculation of Riska and Brown \cite{Riska1}. The couplings to the $\rho N$ and $\omega N$ channels
of the baryon resonances obtained in the hadronic theory
of Ref. \cite{Lutz1}  are rather
different from those obtained in the quark model
of Ref. \cite{Riska1}. They are in general substantially weaker. This issue is
discussed extensively in Ref. \cite{Lutz1}. Consequently the $\pi^-p \rightarrow e^+e^- n$
and $\pi^+n\rightarrow e^+e^- p$ cross sections are
much larger in the calculation of Titov
and K\"{a}mpfer but some trends are similar in both
descriptions. At $\sqrt s$=1.6 GeV, the $\pi^+n\rightarrow e^+e^- p$ cross section
is significantly larger than the $\pi^-p\rightarrow e^+e^- n$ cross section.
This reflects a constructive $\rho^0-\omega$ interference in the first case
and a destructive $\rho^0-\omega$ interference in the latter process as
in our approach.
Much of the strength is provided by the S$_{11}$ resonances. At $\sqrt s$=1.8 GeV,
the cross sections for both reactions are comparable and dominated by the $\omega$-contribution.
The sensitivity of the $\pi^-p \rightarrow e^+e^- n$
and $\pi^+n\rightarrow e^+e^- p$ cross sections to the strength of the
transition couplings of baryon
resonances to vector fields is the key result of both calculations.
The magnitude of these cross sections below threshold would therefore provide very
valuable information on vector meson-nucleon dynamics.
The sensitivity
of the $\pi^-p \rightarrow e^+e^- n$ cross section to the couplings
of baryon resonances to vector meson-nucleon channels can also be seen
from a previous computation \cite{Soyeur} we did in the same coupled-channel framework
as the present calculation.
In Ref. \cite{Soyeur}, the $\pi^-p \rightarrow \rho^0 n$ and $\pi^-p \rightarrow \omega n$
amplitudes were obtained without systematic constraints from meson photoproduction data. This
led to very different couplings, in particular of the N(1520) to the $\rho^0 N$ channel,
and to a constructive rather than a destructive $\rho^0-\omega$ interference.

\newpage
\begin{figure}[ht]
\begin{center}
\mbox{\epsfig{file=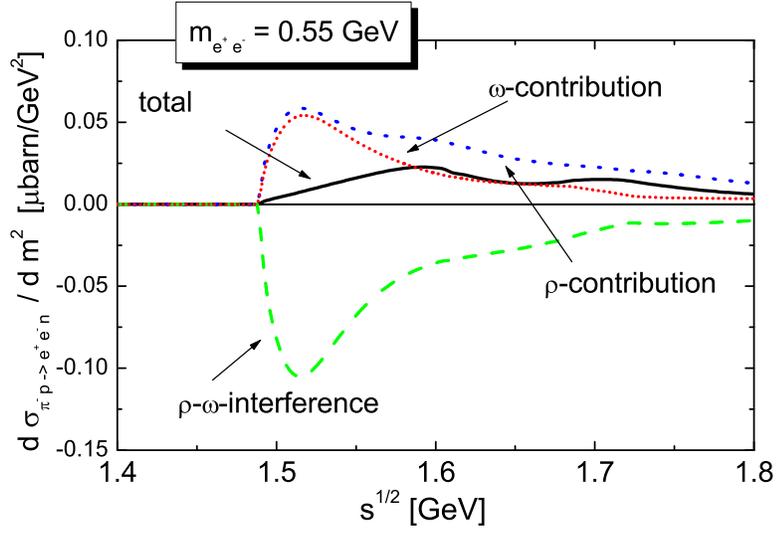, height= 8.2cm}}
\end{center}
\caption{Differential cross section for the $\pi^-p \rightarrow e^+e^- n$
reaction as function of the total pion-nucleon center of mass energy $\sqrt s$
for $e^+e^-$ pairs of invariant mass m$_{e^+e^-}$=0.55 GeV.}
\vskip 0.2truecm
\label{f19}
\end{figure}
\begin{figure}[h]
\begin{center}
\mbox{\epsfig{file=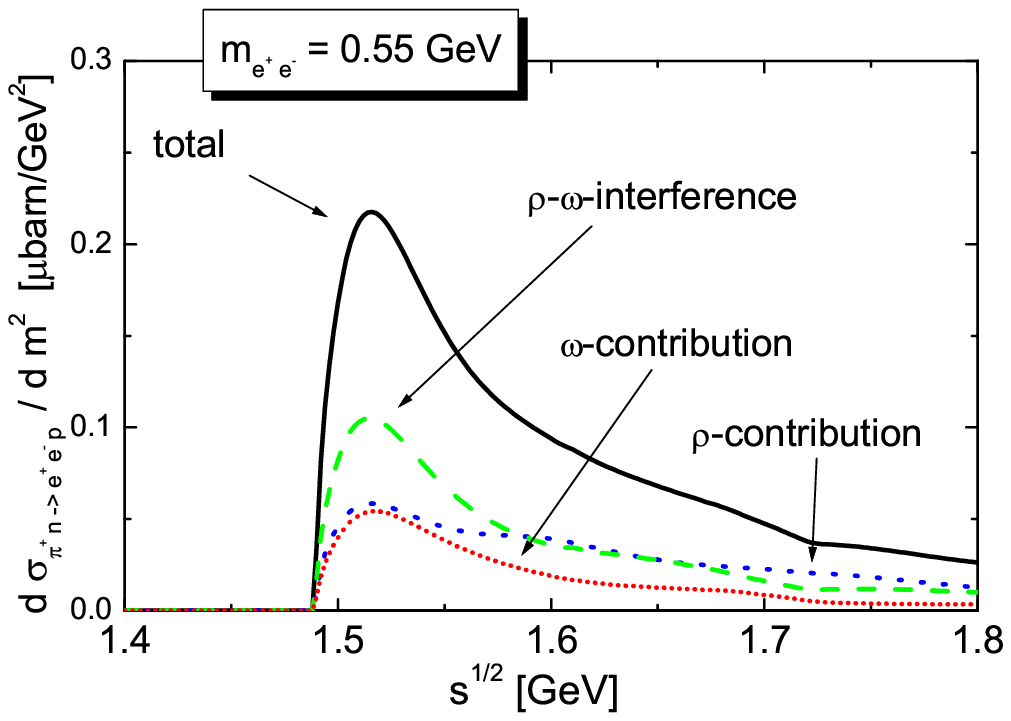, height= 8.2cm}}
\end{center}
\caption{Differential cross section for the $\pi^+n\rightarrow e^+e^- p$
reaction as function of the total pion-nucleon center of mass energy $\sqrt s$
for $e^+e^-$ pairs of invariant mass m$_{e^+e^-}$=0.55 GeV.}
\label{f20}
\end{figure}
\par
\newpage
\begin{figure}[ht]
\begin{center}
\mbox{\epsfig{file=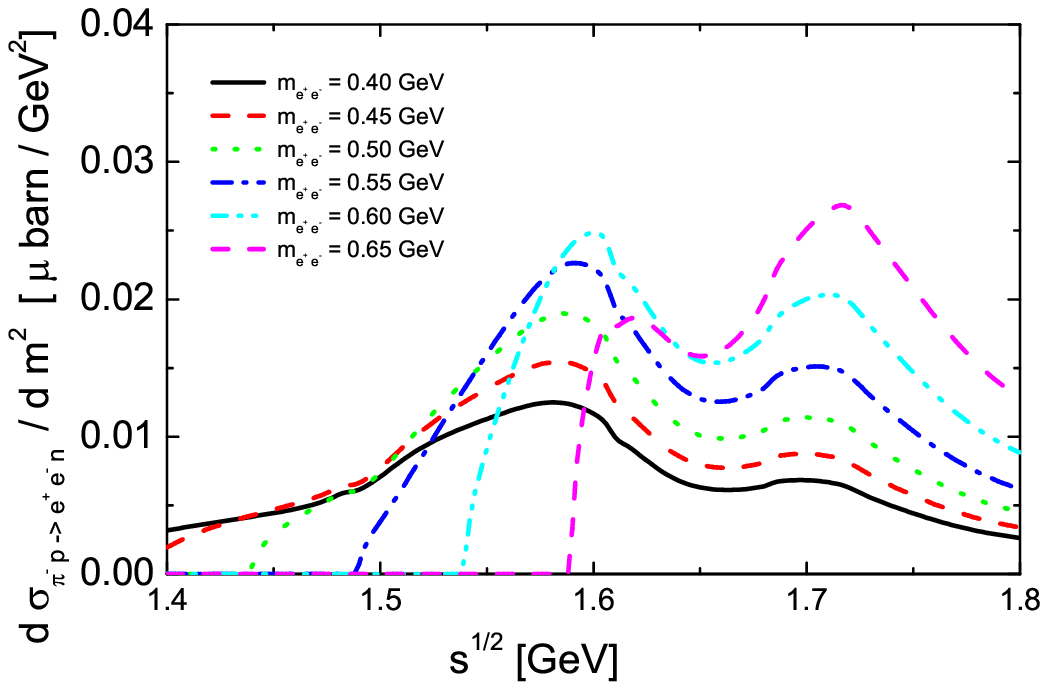, height= 8.5cm}}
\end{center}
\caption{Differential cross section for the $\pi^-p \rightarrow e^+e^- n$
reaction as function of the total pion-nucleon center of mass energy $\sqrt s$
for $e^+e^-$ pairs of invariant masses ranging from 0.40 until 0.65 GeV.}
\label{f21}
\end{figure}
\begin{figure}[h]
\begin{center}
\mbox{\epsfig{file=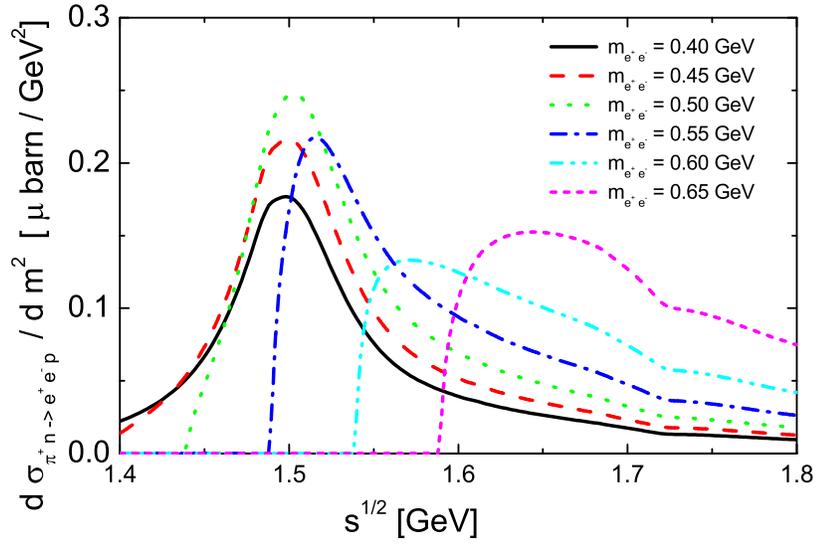, height= 8.5cm}}
\end{center}
\caption{Differential cross section for the $\pi^+n\rightarrow e^+e^- p$
reaction as function of the total pion-nucleon center of mass energy $\sqrt s$
for $e^+e^-$ pairs of invariant masses ranging from 0.40 until 0.65 GeV.}
\label{f22}
\end{figure}
\par
\newpage
\section{Conclusion}

We have computed the $e^+e^-$ pair invariant mass distributions for the $\pi^-p \rightarrow e^+e^- n$
and $\pi^+n\rightarrow e^+e^- p$ reactions below and close to the vector meson production threshold
($\sqrt s$=1.72 GeV).

We employ the $\pi N\rightarrow\rho^0 N$ and $\pi N\rightarrow\omega N$ amplitudes
obtained in a recent relativistic and unitary coupled-channel approach to
meson-nucleon scattering \cite{Lutz1}. This description reproduces a large
body of data on pion-nucleon elastic and inelastic scattering and on meson photoproduction off nucleons
in the energy range $1.4<\sqrt s <1.8$ GeV. In the model, pion-nucleon
resonances are generated dynamically and the coupling strengths of these resonances to vector meson-nucleon
channels are predicted. These couplings are not well-known.
% and are most directly accessible through $e^+e^-$ decays.

Using the Vector Meson Dominance assumption, we have shown that the differential cross sections for
the $\pi^-p \rightarrow e^+e^- n$
and $\pi^+n\rightarrow e^+e^- p$ reactions below the $\omega$-threshold are very sensitive to the
coupling of low-lying baryon resonances to vector meson-nucleon
final states. We find that the $\rho^0-\omega$ interference is destructive in the
$\pi^-p \rightarrow e^+e^- n$ channel and constructive in the $\pi^+n\rightarrow e^+e^- p$ channel.
We predict a very small cross section for the $\pi^-p \rightarrow e^+e^- n$ reaction
below threshold and a sizeable cross section for the $\pi^+n\rightarrow e^+e^- p$ reaction
in this energy range. Above the $\omega$-meson production threshold, both cross sections are
comparable and much larger.

The magnitude of the $\pi^-p \rightarrow e^+e^- n$
and $\pi^+n\rightarrow e^+e^- p$  differential cross sections below the $\omega$-threshold depends strongly
on the structure and dyna\-mics of baryon resonances. These reactions deserve
experimental studies. Such a programme could be carried at GSI (Darmstadt) using
the available pion beam and the HADES spectrometer \cite{HADES}. These measurements
would provide a necessary step towards the understanding of $e^+e^-$ pair
production in pion-nucleus reactions and
in general significant constraints on the propagation of vector mesons in the
nuclear medium.

\end{document}